# Energy-composition relations in Ni$_3$(Al$_{1-x}$X$_x$) phases

Nikolai A. Zarkevich,[1] Timothy M. Smith,[2] John W. Lawson[1]

[1] Intelligent Systems Division, NASA Ames Research Center, Moffett Field, CA 94035, USA
[2] NASA Glenn Research Center, 21000 Brook Park Rd., Cleveland, OH 44135, USA

**Abstract**: The secondary phase, such as Ni$_3$Al-based L1$_2$ γ', is crucially important for precipitation strengthening of superalloys. Composition-structure-property relations provide useful insights for guided alloy design. Here we use density functional theory combined with the multiple scattering theory to compute dependencies of the structural energies and equilibrium volumes versus composition for ternary Ni$_3$(Al$_{1-x}$X$_x$) alloys with X=(Ti, Zr, Hf; V, Nb, Ta; Cr, Mo, W) in L1$_2$, D0$_{24}$, and D0$_{19}$ phases with a homogeneous chemical disorder on the (Al$_{1-x}$X$_x$) sublattice. Our results provide a better understanding of the physics in Ni$_3$Al-based precipitates and facilitate design of next-generation nickel superalloys with precipitation strengthening.

## 1. Introduction:

Computed dependencies of the relative structural energies on composition can be used to design multi-phase materials and alloys with improved thermomechanical properties (in particular, with better creep and higher strength) [1,2]. Recently it was shown [3] that local phase transformations inside Ni$_3$Al-based precipitates improve creep at the elevated operation temperatures *T* in Ni-based superalloys used in jet engines. At high *T*, local phase transformations are assisted by atomic diffusion. Structural defects (such as stacking faults) interact with diffusing solute atoms and are stabilized by the local chemical composition. Attractive interactions result in energy reduction after diffusion of particular chemical elements towards defects, which act as sinks. Stabilization of stacking faults inside L1$_2$ precipitates reduces creep and improves mechanical strength of Ni superalloys [4,5]. The stacking of atomic layers in intrinsic and extrinsic stacking faults in the L1$_2$ phase locally looks like D0$_{19}$ and D0$_{24}$ structures, respectively. Consequently, the stacking fault energy correlates with the energy difference between the relevant structures. To provide guidance for alloy design, we computed the compositional dependences of these structural energies.

The relevant atomic structures are compared in Figure 1. They differ by the stacking of the atomic layers along the cubic [111] (in L1$_2$) and hexagonal (hex) [0001] directions (in D0$_{19}$ and D0$_{24}$ structures). The periodic stacking sequences are [AB] in D0$_{19}$ and [ABAC] in D0$_{24}$; both structures can be viewed locally as stacking faults within the [ABC] stacking in the cubic L1$_2$ structure along the [111] direction. In Ni$_3$(Al$_{1-x}$X$_x$) alloys with chemical disorder on the (Al$_{1-x}$X$_x$) sublattice, this sublattice is simple cubic in L1$_2$, hexagonal close-packed (hcp) in D0$_{19}$, and double hcp (dhcp) in D0$_{24}$.

Novelty of our results consists in providing the previously unknown composition-property relations, shown in Figs. 2 and 3. To compute the energy-composition dependencies from the first principles, we use the well-established theoretical methods [6]. The computed structural formation energies and equilibrium volumes (per atom) versus composition are shown in Figure 2. Relative energies versus composition are in Figure 3. Known composition-property dependencies allow to improve alloys by compositional adjustment.

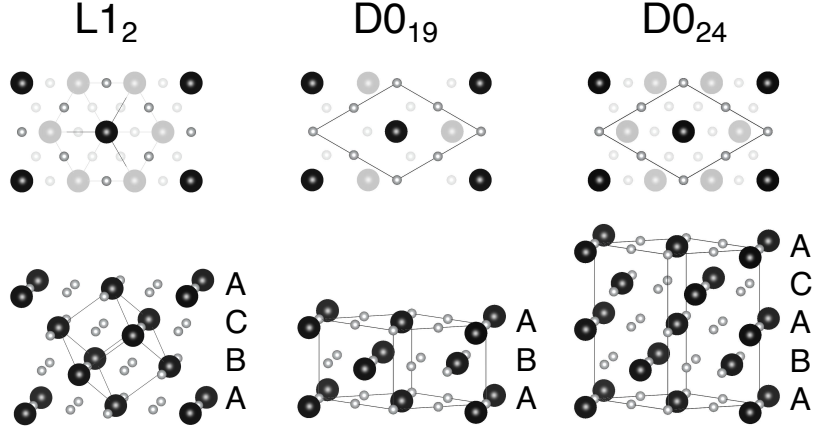

Fig. 1: Atomic structures of L1$_2$ ($\gamma'$), D0$_{19}$ ($\chi$), and D0$_{24}$ ($\eta$) phases, shown for fully relaxed Ni$_3$Ti. The top panel shows cubic (111) and hexagonal (0001) projections, with shaded lower layers. The lower panel compares stacking of atomic layers. Ti is large black; Ni is small grey.

## 2. Methods:

We combine density functional theory (DFT) with multiple scattering theory (MST) to compute sets of energies [7] of crystal structures with chemical disorder on sublattices. The homogeneous atomic disorder without a short-range order (SRO) is considered within the coherent potential approximation (CPA) [8] in the Korringa-Kohn-Rostocker (KKR) formalism [9,10]. The terminal binary Ni$_3$X structures are addressed by both KKR and full-potential DFT methods; the latter provides higher accuracy for structural formation energies, determined relative to the ground states of elemental solids. To get the advantages of both full-potential and MST methods, we use the full-potential formation energies of the terminal Ni$_3$X structures (see Table 1) and combine them with the KKR-CPA mixing energies of disordered Ni$_3$(Al$_{1-x}$X$_x$) structures. The mixing energies are defined relative to those of the terminal Ni$_3$X structures in the same phase. All equilibrium energies and volumes are computed at zero pressure and zero temperature $T = 0$ K; values are reported per atom, unless specified otherwise.

We use the all-electron KKR-CPA code [11] to find mixing energies of Ni$_3$(Al$_{1-x}$X$_x$)$_1$ alloys with a homogeneous chemical disorder on the (Al$_{1-x}$X$_x$) sublattice. We use the full-potential VASP code [12] to compute formation energies of the fully ordered binary structures at the terminal Ni$_3$X compositions. In both codes we use the PBEsol [13] exchange correlation functional (XC=116133).

The KKR-CPA spin-polarized calculations [8-10] were performed in primitive unit cells. We used two $k$-point meshes for the Brillouin Zone (BZ) integration. The primary (secondary) $k$-mesh was 12×12×12 (8×8×8) for the cubic L1$_2$, 8×8×4 (6×6×4) for the hexagonal (hex) D0$_{24}$ with fixed $c/a=(8/3)^{1/2}$ (ideal), and 8×8×10 (6×6×6) for the hexagonal D0$_{19}$ with fixed $c/a=(2/3)^{1/2}$ (ideal). We included $s$, $p$, $d$, and $f$ orbitals ($l_{max}=3$) in the basis inside the atomic spheres. For contour integration in the complex plane, we fixed the bottom energy at or below $E_{bot}=-0.9$ Ry. We used the muffin-tin approximation with periodic boundary corrections. At each composition $x$, the equilibrium volume $V_0$ and the minimal energy $E_0$ were found by fitting the Birch-Murnaghan [14,15] equation of state (EOS) defining the energy versus volume $E(V)$ relation to 5 DFT points (N$_{eos}$=5) with 1.5% step in the lattice constant $a$. To check accuracy, we used the fitted linear $V_0(x)$ dependence in each Ni$_3$(Al$_{1-x}$Hf$_x$) phase and directly computed DFT energies $E[V_0(x)]$, which agreed with the EOS energies $E_0(x)$ within the DFT error bars.

The VASP code [12] was compiled with the C2NEB subroutine [16,17]. A dense Γ-centered Monkhorst-Pack mesh [18] with ≥60 k-points per inverse Angstrom (Å) was used for the BZ integration. The plane-wave energy cutoff was increased to ENCUT=650 eV. We used Gaussian smearing (ISMEAR=0) with SIGMA=0.043 eV, corresponding to $k_B T$ at $T$=500 K, where $k_B$ is the Boltzmann constant. DFT energy was obtained by extrapolation to zero smearing. Stacking fault energies were computed in a supercell with 40Å between the periodic stacking faults; the energy of an ideal crystal was computed in a primitive unit cell.

## 3. Results:

Using density functional theory[19-21] and multiple scattering theory[6,22], we computed the equations of state, formation and mixing energies of the ordered $Ni_3X$ and disordered $Ni_3(Al_{1-x}X_x)$ alloys with X=(Ti, Zr, Hf; V, Nb, Ta; Cr, Mo, W) in $L1_2$, $D0_{19}$, and $D0_{24}$ phases, shown in Fig. 1. Formation energies (computed using VASP, relative to the ground states of elemental solids) of the terminal $Ni_3X$ compositions are in Table 1. Mixing energies (relative to the ground states of the terminal $Ni_3Al$ and $Ni_3X$ compositions) were computed using KKR-CPA. Formation energies versus composition for the partially disordered $Ni_3(Al_{1-x}X_x)$ structures (with a homogeneous disorder on the Al+X sublattice) are shown in Fig. 2. The EOS equilibrium energies $E_0$ and volumes $V_0$ are shown as dots. For X=(Hf, Ta) in $Ni_3(Al_{1-x}Hf_x)$ and $Ni_3(Al_{1-x}Ta_x)$, the $E(x)$ lines are the 6th degree polynomials, fitted to the EOS data points $E_0$. For X=(Cr, W), the $V(x)$ lines are the 4th degree polynomials from the least-squares fit to the EOS volumes $V_0$. For the other elements X, lines connect the computed EOS energies $E_0$ and volumes $V_0$, shown as dots.

For completeness, we also provide relative energies $\Delta E$ in Fig. 3, which is complementary to Fig. 2. The differences $\Delta E$ are defined relative to $E_0(L1_2)$ at the same composition $x$. These differences $\Delta E(D0_{24} - L1_2) = E(D0_{24}) - E(L1_2)$ and $\Delta E(D0_{19} - L1_2) = E(D0_{19}) - E(L1_2)$ in Fig. 3 provide important additional information, while the shape of each $E(x)$ curve in Fig. 2 is related to the stability of each phase. The energy differences $\Delta E$ alone are not sufficient for materials design, especially if an unstable phase at some composition $x$ transforms to another phase or segregates into other compositions. For example, $E_0(x)$ curves for all 3 phases of $Ni_3(Al_{1-x}Zr_x)_1$ alloys at 0<$x$<0.8 point at a tendency to segregate into Zr-rich and Zr-deficient components; for a concave $E(x)$ dependency with a negative curvature such segregation would result in energy lowering [23].

Our first-principles calculations predict a change of the lowest-energy phase from $L1_2$ $Ni_3Al$ to $D0_{24}$ for group 4 elements X=(Ti, Zr, Hf) and to $D0_{19}$ for group 5 elements X=(V, Nb, Ta), with a possible intermediate $D0_{24}$ phase for X=(V, Nb). However, we did not consider $D0_{22}$ and $D0_a$ phases of $Ni_3(V,Nb)$ [24] and $Ni_3Ta$ [25], reported as more stable than $D0_{19}$ in the phase diagrams [26,27].

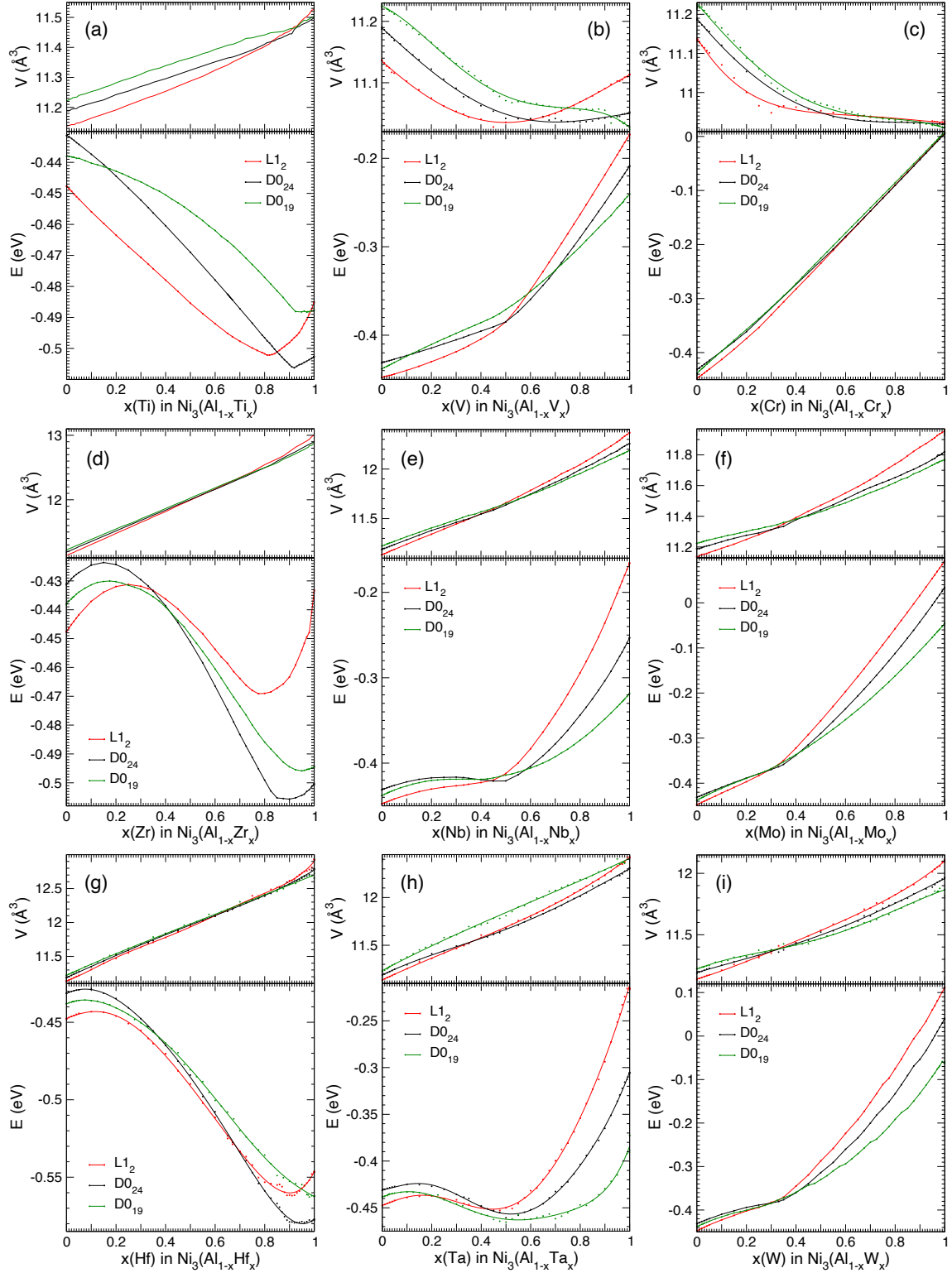

Fig. 2: Equilibrium volume $V$ [Å$^3$/atom] and formation energy $E$ [eV/atom] of the Ni$_3$(Al$_{1-x}$X$_x$) phases with homogeneous atomic disorder on the Al/X sublattice for group 4, 5, and 6 elements X=(Ti, Zr, Hf; V, Nb, Ta; Cr, Mo, W). The terminal formation energies of the binary Ni$_3$X alloys are from the full-potential VASP, while the relative mixing energies are from the KKR-CPA calculations. L1$_2$ is red, D0$_{24}$ is black, D0$_{19}$ is green.

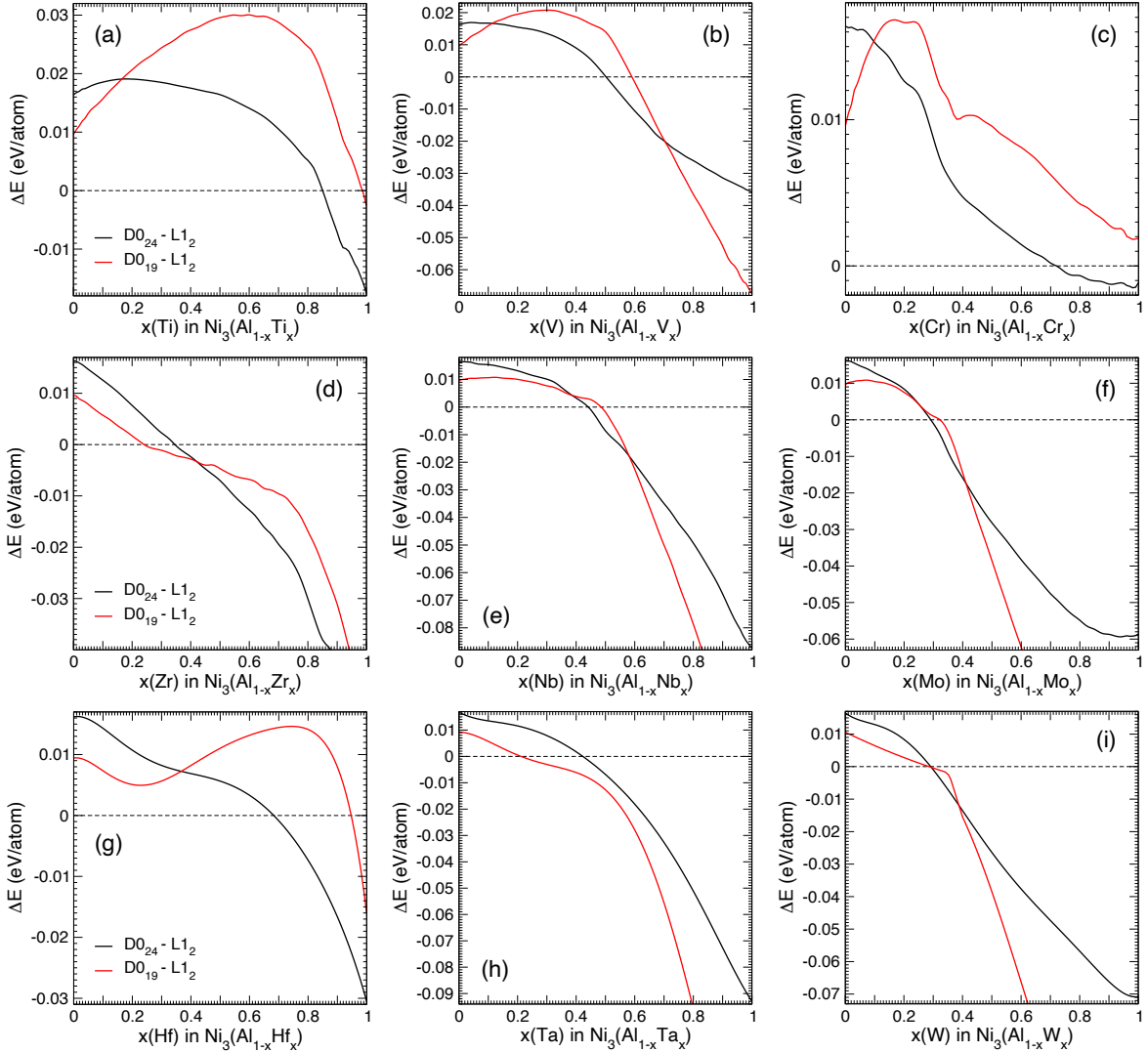

Fig. 3: Relative energies [eV/atom] of the $Ni_3(Al_{1-x}X_x)$ phases. Data in Fig. 2 was approximated by cubic splines and $E_0(L1_2)$ was subtracted. $\Delta E(D0_{24} - L1_2)$ is black and $\Delta E(D0_{19} - L1_2)$ is red.

The equation of state contains equilibrium energy $E_0$, volume $V_0$, bulk modulus $B_0$, and its dimensionless pressure derivative $B'_0 = \left(\frac{\partial B}{\partial P}\right)_T$ at constant temperature $T$. The computed values of $B_0$ and $B'_0$ in Table 2 were assessed from the EOS fitted to DFT KKR data. For the computed compositional dependences of $B_0$ and $B'_0$ we found that the differences between the linear and higher-degree polynomial approximations were within the error bars. Directly computed values of $B_0$ for $Ni_3Al$ and $Ni_3Ti$ in Table 2 can be compared with those at the terminal compositions from a linear fit of $B_0(x)$ for the $Ni_3(Al_{1-x}Ti_x)$ system: $B_0(L1_2, x=0) = 193.13$ GPa ≈ 193.1 GPa = $B_0(L1_2, Ni_3Al)$; $B_0(L1_2, x=1) = 203.49$ GPa ≈ 203.1 GPa = $B_0(L1_2, Ni_3Ti)$; $B_0(D0_{24}, x=0) = 193.16$ GPa ≈ 193.8 GPa = $B_0(D0_{24}, Ni_3Al)$; $B_0$ $(D0_{24}, x=1) = 204.92$ GPa ≈ 204.7 GPa = $B_0(D0_{24}, Ni_3Ti)$. The computed bulk modulus $B_0$ of $L1_2$ $Ni_3Al$ is 193 GPa, and the experimental measurements range from 171 GPA [28] and 173.9 GPa [29] to 229.2 GPa [30]. We conclude that our first-principles results reasonably agree with the available experimental data.

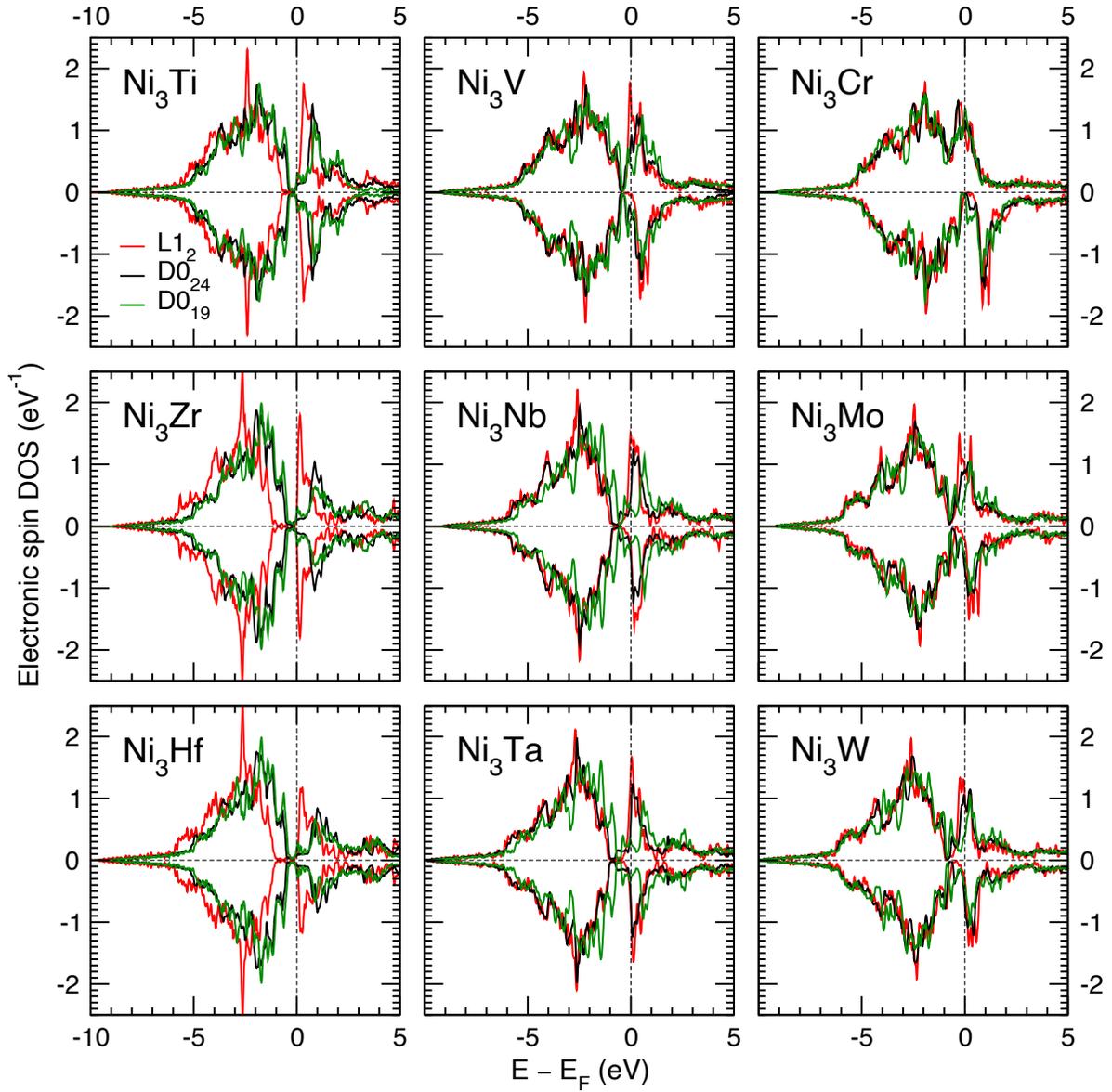

Fig. 4. Spin DOS in L1$_2$, D0$_{24}$ and D0$_{24}$ phases of Ni$_3$X compounds.

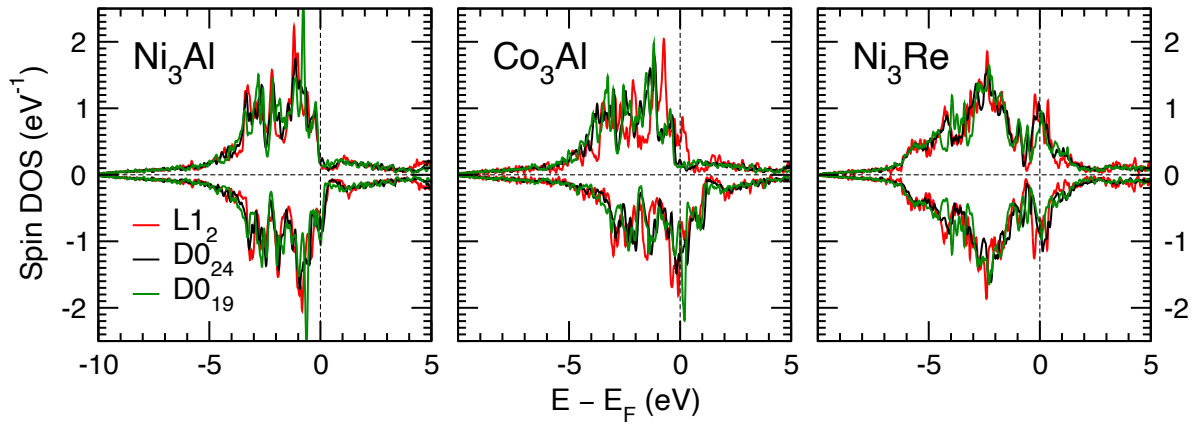

Fig. 5. Spin DOS in L1$_2$, D0$_{24}$ and D0$_{24}$ phases of Ni$_3$Al, Co$_3$Al, and Ni$_3$Re compounds.

Table 1. The computed formation energies [eV/atom] and volumes [Å$^3$/atom] of Ni$_3$X compounds with L1$_2$, D0$_{19}$, and D0$_{24}$ structures, fully relaxed in VASP.

| Ni$_3$X | E (eV/atom) | | | V (Å$^3$/atom) | | |
|---|---|---|---|---|---|---|
| | L1$_2$ | D0$_{24}$ | D0$_{19}$ | L1$_2$ | D0$_{24}$ | D0$_{19}$ |
| Ni$_3$Al | -.4477 | -.4313 | -.4381 | 10.90 | 10.91 | 10.92 |
| Ni$_3$Ti | -.4852 | -.5028 | -.4878 | 11.25 | 11.24 | 11.25 |
| Ni$_3$Zr | -.4332 | -.5007 | -.4943 | 12.63 | 12.62 | 12.62 |
| Ni$_3$Hf | -.5465 | -.5774 | -.5621 | 12.37 | 12.34 | 12.35 |
| Ni$_3$V | -.1729 | -.2088 | -.2404 | 10.81 | 10.77 | 10.73 |
| Ni$_3$Nb | -.1658 | -.2531 | -.3184 | 11.99 | 11.92 | 11.88 |
| Ni$_3$Ta | -.2155 | -.3054 | -.3727 | 11.96 | 11.88 | 11.83 |
| Ni$_3$Cr | +.0062 | +.0051 | +.0081 | 10.66 | 10.64 | 10.60 |
| Ni$_3$Mo | +.0924 | +.0338 | -.0460 | 11.59 | 11.49 | 11.46 |
| Ni$_3$W | +.1124 | +.0403 | -.0593 | 11.62 | 11.53 | 11.49 |

Table 2. Computed EOS parameters: bulk modulus $B_0$ (GPa) and $B'_0$ (dimensionless) from KKR.

| Ni$_3$X | $B_0$ (GPa) | | | $B'_0$ | | |
|---|---|---|---|---|---|---|
| | L1$_2$ | D0$_{24}$ | D0$_{19}$ | L1$_2$ | D0$_{24}$ | D0$_{19}$ |
| Ni$_3$Al | 193.1 | 193.8 | 193.7 | 4.46 | 4.42 | 4.62 |
| Ni$_3$Ti | 203.1 | 204.7 | 204.6 | 4.51 | 4.47 | 4.36 |
| Ni$_3$Zr | 172.5 | 177.8 | 178.1 | 4.32 | 4.15 | 4.11 |
| Ni$_3$Hf | 182.8 | 175.0 | 192.5 | 4.36 | 4.27 | 4.23 |
| Ni$_3$V | 220.6 | 226.0 | 228.3 | 4.55 | 4.72 | 4.53 |
| Ni$_3$Nb | 207.0 | 213.1 | 217.7 | 4.32 | 4.42 | 4.49 |
| Ni$_3$Ta | 217.8 | 224.4 | 225.1 | 4.33 | 4.44 | 5 |
| Ni$_3$Cr | 214.6 | 216.4 | 217.8 | 5.02 | 4.96 | 5.04 |
| Ni$_3$Mo | 226.7 | 229.0 | ~238 | 4.42 | 4.41 | 4.47 |
| Ni$_3$W | 237.5 | 248.7 | ~270 | 4.56 | 4.20 | 4.2 |

For changing composition $x$, magnetic and electronic structure changes in the Ni$_3$(Al$_{1-x}$Ti$_x$)$_1$ system [31]. Ni$_3$Al is magnetic, while Ni$_3$Ti is non-magnetic, with zero atomic magnetic moments, see Fig. 4 in Ref. [31]. The rapid change of the electronic density of states at the Fermi energy $E_F$ is responsible for the kink of the $E(x)$ curve in Ni$_3$(Al$_{1-x}$Ti$_x$)$_1$, present in each of the 3 phases. $E_F$ is in the pseudo-gap in Ni$_3$Ti, but not in Ni$_3$Al, see Figs. 4 and 5. Ni$_3$(Al$_{1-x}$Ti$_x$)$_1$ crystal structure changes from L1$_2$ at smaller $x$ to D0$_{24}$ at $x \geq 0.875$. Similar changes of electronic, magnetic, and atomic structure are also expected in other Ni$_3$(Al$_{1-x}$X$_x$)$_1$ systems.

The computed equation of state can be affected by a magnetic phase transition. This is the case, for example, in D0$_{19}$ Ni$_3$V: due to the disappearance of atomic magnetic moments near $V_0$, the EOS fit of both magnetic states at $V>V_0$ and non-magnetic states at $V<V_0$ results in a larger fitting error and less accurate value of $B'_0$. Two EOS fits of either non-magnetic states at $V<V_0$ (shown in Table 2) or magnetic states at $V>V_0$ provide two different equations of state with similar $E_0$, but different $B_0$ and $B'_0$ values.

Our attempts to fit the EOS for the D0$_{19}$ (not the ground state) structure in Ni$_3$Mo and Ni$_3$W resulted in a large fitting error, originating from peculiarities of electronic structure. The equilibrium values of $E_0$ and $V_0$ were found reliably by energy minimization. However, the EOS fit of the higher-order terms $B_0$ and $B'_0$ was noisy: slightly different calculations provided different results. Ni$_3$Mo crystallized below 910°C (1183 K) in the orthorhombic D0$_a$ oP8 β-Cu$_3$Ti structure with *Pmmn* space group (no. 59) [32], which was claimed to be stable [33]. Stability of Ni$_3$W oP8 structure [34] is debatable [35] and can be influenced by carbon [36]. Unstable $Ni_3W$ $L1_2$ and $D0_{24}$ structures have positive formation energies (see Table 1 and Fig. 2), while $Ni_3W$ $D0_{19}$ structure is not observed experimentally [37]. For comparison, the Cr-Ni system segregates into Cr and Ni solid solutions [38]; there are no stable $Ni_3Cr$ compounds, in agreement with our calculations.

Fig. 2 allows to roughly estimate temperature-dependent solubility limits of dopants in the $Ni_3Al$ $L1_2$ phase, using a rapid design estimate of phase-segregation temperature [39]. However, consideration of other phases, such as the orthorhombic $Ni_3(Nb_{0.8}Ti_{0.2})$ δ phase [40], is beyond the scope of present work.

The Ti-rich $Ni_3(Al_{1-x}Ti_x)_1$ alloy was claimed [31] to be a compositional glass – an analogue of a spin glass in the compositional space. Such systems can be described by a truncated cluster expansion [41] with a degeneracy among the interactions [42] and frustration of the ground states [43]. At certain compositions $x$, we expect similarly frustrated ground states in several $Ni_3(Al_{1-x}X_x)_1$ systems.

Interestingly, in the $Ni_3(Al_{1-x}X_x)_1$ systems with X=(Ti, V, Nb), the full-potential VASP calculations predict repulsion of X=(Ti, V, Nb) from the stacking fault at small concentration $x$ and attraction to the stacking fault at larger $x$. For $Ni_3(Al_{1-x}Nb_x)_1$, we checked this for both intrinsic and extrinsic stacking faults. The KKR-CPA results in Fig. 3 indicate attraction of Nb to the stacking fault at larger $x$; however, repulsion of Nb at small concentrations $x$ from both intrinsic and extrinsic stacking faults was unexpected [44].

In $L1_2$ $Ni_3Nb$ (at $x=1$), the computed stacking fault energies are negative for both intrinsic and extrinsic stacking faults; this points at instability of $L1_2$ $Ni_3Nb$ structure. Indeed, both $D0_{24}$ and $D0_{19}$ structures are lower in energy than $L1_2$. At some composition $x$, a stacking fault energy changes its sign (i.e., becomes zero) in the $Ni_3(Al_{1-x}Nb_x)_1$ system.

In $Ni_3Al$ (at $x=0$), our computed energy of the intrinsic stacking fault is 0.054 $J/m^2$; this value reasonably agrees with those ranging from 0.037 to 0.092 $J/m^2$ in the literature [44-51]. Depending on the distance $L$ between the periodic stacking faults in a supercell, the intrinsic stacking fault energy varies from 0.065 $J/m^2$ at $L≈10$ Å to 0.054 $J/m^2$ at $L>40$ Å; a monotonic decrease of energy with distance at $L>10$ Å points at repulsion between the stacking faults.

Stabilization of stacking faults in $L1_2$ precipitates reduces creep of Ni-based superalloys at high operating temperatures [3-5]. Local phase transformation at stacking faults with increased concentration of dopants impedes propagation of additional dislocations along the stacking fault and thus improves creep resistance. The $D0_{24}$ and $D0_{19}$ structures locally look like $L1_2$ with periodic stacking faults, and the energy differences in Fig. 3 correlate with the stacking fault energies. Thus, our results allow the engineering of stacking fault energies and improved creep in Ni superalloys by compositional adjustment. Discussion of a predictor for choosing the right chemical elements with appropriate concentrations that promote formation and stabilization of η (χ) phases along the superlattice extrinsic (intrinsic) stacking faults can be found in the literature [3-5]. A limited amount of relevant DFT data can be found in Fig. 6 in Ref. 3. Here in Figs. 2 and 3 we show that the energy-composition relations are nonlinear. This detailed information can be used as an improved predictor, which takes into account the non-linearity of relative energies versus composition.

## 4. Discussion:

For completeness, we discuss superalloys and their precipitation strengthening. A superalloy is a high-performance metallic alloy that is capable to operate at high temperatures – a fraction of melting point [2,52,53]. Superalloys have common characteristics, retained at operating temperature, such as mechanical strength, low creep, surface stability, corrosion and oxidation resistance, radiation tolerance, and metallic electrical and thermal conductivity [54]. Due to their properties, superalloys are used in load-bearing structures at high temperatures and stresses, in highly corrosive or radioactive environments: in engines, generators, heat exchanges, and nuclear reactors. Superalloys find applications in energy, nuclear, chemical processing, automotive, marine, and airspace industries [52,53]. Both superalloys and multiprincipal-element alloys are suitable for 3D printing [55].

Nickel-based superalloys are known for many decades [56]. They are composed by the Ni-rich fcc solid solition ($\gamma$ phase) and precipitates, most of which have the cubic $L1_2$ cP4 crystal structure with $Cu_3Al$ prototype ($Ni_3Al$ $\gamma$' phase) [57]. The competing phases include the tetragonal $D0_{22}$ ($Ni_3Nb$) $\gamma$" [24], hexagonal $D0_{24}$ ($Ni_3Ti$) $\eta$ [31], hexagonal $D0_{19}$ ($Ni_3Sn$-type) $\chi$, orthorhombic $D0_a$ ($Cu_3Ti$-type) $Ni_3Ta$ [25], carbides, etc.

The local phase transformations inside $\gamma$' precipitates followed by diffusion-driven stabilization of the stacking faults were shown to improve mechanical strength and creep in Ni superalloys [4,5]. The solute atoms interact with defects [58], such as dislocations [59], stacking faults [5], twins [60], grain boundaries [61], and surfaces [1]. Altered chemical composition near the surface affects corrosion and oxidation resistance. Stabilization of grain boundaries may improve strength of a polycrystal towards that of a single crystal.

Previously, the interaction energies between solutes (Co, Cr, Nb, Ta) and stacking faults were computed [4]. We find nonlinearity of the key energies versus composition, see Figs. 2 and 3. One reason for that is a change of the electronic structure from that with a relatively high density of states at the Fermi energy in $Ni_3Al$ (Fig. 5) to a low one (with the Fermi energy in the pseudo-gap, for example, in $Ni_3Ti$, see Fig. 4). The computed from the first principles nonlinear (and sometimes non-monotonic) property-composition relations constitute the main novelty of our present research.

**5. Summary**:

We have computed the compositional dependencies of the equations of state, structural energies and equilibrium volumes (per atom) for ternary $Ni_3(Al_{1-x}X_x)$ alloys with X=(Ti, Zr, Hf; V, Nb, Ta; Cr, Mo, W) in $L1_2$, $D0_{24}$, and $D0_{19}$ phases with atomic disorder on the $(Al_{1-x}X_x)$ sublattice. We considered both formation and relative energies and found their nonlinear dependencies versus composition. Our results provide better understanding of precipitation in multicomponent Ni superalloys. Our *ab initio* data is used for designing next-generation alloys with improved properties. [55]

**Acknowledgements**: We acknowledge funding by NASA's Aeronautics Research Mission Directorate (ARMD) via Transformational Tools and Technologies (TTT) Project. We thank Anupa R. Bajwa, Mikhail Mendelev, Valery V. Borovikov, and Shreyas J. Honrao for discussion.

**Author contributions**: Nikolai Zarkevich conceived the idea, performed calculations, and wrote the manuscript. Timothy Smith and John Lawson participated in discussions and editing.

**Conflicts of Interest:** The authors declare no conflict of interest. Opinion of the authors does not represent opinions of any governmental organizations.